\documentclass[openacc]{rstransa}

\begin{document}

\title{John Couch Adams: mathematical astronomer, college friend of George Gabriel Stokes and promotor of women in astronomy}

\author{
Davor Krajnovi\'c$^{1}$}

\address{$^{1}$Leibniz-Institut f\"ur Astrophysik Potsdam (AIP), An der Sternwarte 16, D-14482 Potsdam, Germany }

\subject{xxxxx, xxxxx, xxxx}


\corres{Davor Krajnovi\'c\\
\email{dkrajnovic@aip.de}}

\begin{abstract}
John Couch Adams predicted the location of Neptune in the sky, calculated the expectation of the change in the mean motion of the Moon due to the Earth's pull, and determined the origin and the orbit of the Leonids meteor shower which had puzzled astronomers for almost a thousand years. With his achievements Adams can be compared with his good friend George Stokes. Not only were they born in the same year, but were also both senior wranglers, received the Smith's Prizes and Copley medals, lived, thought and researched at Pembroke College, and shared an appreciation of Newton. On the other hand, Adams' prediction of Neptune's location had absolutely no influence on its discovery in Berlin. His lunar theory did not offer a physical explanation for the Moon's motion. The origin of the Leonids was explained by others before him. Adams refused a knighthood and an appointment as Astronomer Royal. He was reluctant and slow to publish, but loved to derive the values of logarithms to 263 decimal places. The maths and calculations at which he so excelled mark one of the high points of celestial mechanics, but are rarely taught nowadays in undergraduate courses. The differences and similarities between Adams and Stokes could not be more striking. This volume attests to the lasting legacy of Stokes' scientific work. What is then Adams' legacy? In this contribution I will outline Adams' life, instances when Stokes' and Adams' lives touched the most, his scientific achievements and a usually overlooked legacy: female higher education and support of a woman astronomer.

\end{abstract}



\maketitle

\section{Adams, the celebrity}

The opening sentences of John Couch Adams' obituary boldly states that "England has lost the greatest mathematical astronomer she has ever produced, Newton alone excepted" \cite{1892Obs....15..173G}. Adams achieved a true celebrity  status early in his life and enjoyed a long and fruitful career. His scientific achievements were praised, respected and rewarded, first with the Royal Society Copley Medal and later with the Gold Medal of the Royal Astronomical Society. He was offered a knighthood and received honorary degrees from the Universities of Oxford, Dublin, Bologna and Cambridge. His professional status was never in question. He obtained a fellowship at St. John's College immediately after undergraduate studies and had little problem finding a new fellowship at Pembroke College after the first fellowship expired. His name is associated with a mathematics prize and, after his death, a John Couch Adams Stipend was established for an astronomer. Adams first became the Regius Professor of Mathematics at the University of St. Andrews and soon after that obtained the Lowndean Professorship in Astronomy and Geometry in Cambridge. He became the director of the Cambridge University Observatory and was offered the job of Astronomer Royal. Upon his death Queen Victoria expressed a wish to bury him in Westminster Abbey, next to kings, queens, and other heroes of Britain. 

All this would not have been possible if Adams were anything less than a most brilliant scholar, as well as a person who showed kindness and inspired respect. Perhaps the best description of Adams is given by his physician, Dr. Donald MacAlister: \textit{"His earnest devotion to duty, his simplicity, his perfect selflessness, were to all who knew his life at Cambridge a perpetual lesson, more eloquent than speech. From the time of his first great discovery scientific honours were showered upon him, but they left him as they found him -- modest, gentle and sincere. Controversies raged for a time around his name, national and scientific rivalries were stirred concerning his work and its reception, but he took no part in them, and would generously have yielded to others' claims more than his greatest contemporaries would allow to be just. With a single mind for pure knowledge he pursued his studies, here bringing a whole chaos into cosmic order, there vindicating the supremacy of a natural law beyond the imagined limits of its operation; now tracing and abolishing errors that had crept into the calculations of the acknowledged masters of his craft, and now giving time and strength to resolving the self-made difficulties of a mere beginner, and all the while with so little thought of winning recognition or applause that much of his most perfect work remained for long, or still remains, unpublished"} \cite{Glaisher 1896}.

There is no doubt that upon Adams' death there were many people who missed him as a companion, friend, teacher and a paragon combining the best of human qualities and scientific achievements, who can be compared with the giants of thought of the past, like Newton, Gauss or Laplace (e.g. \cite{Glaisher 1896}). Yet, in the scientific sense, if one removes layers of national pride, Adams today is a poorly known figure. His work did not inspire any new developments in physics or mathematics. Towards the actual discovery of Neptune, which propelled him into the stardom of British science, Adams did not contribute at all \cite{Encke 1846}. In his most influential achievement, on the theory of the Moon's mean motion, Adams showed that the contribution of the diminution of the eccentricity of the Earth's orbit can account for only about half of the observed effect, correcting the work of a number of "geometers" \cite{1853MNRAS..14...59A}, but he did not attempt to pursue the matter further nor provide the full physical explanation. It is simply not possible to organise a meeting such as this one for Adams, where we have seen papers that apply ideas originated by Stokes some 150 years ago to fully modern problems and applications ranging from flows of fluids over corals to plasma confinement in fusion reactors.  

\begin{figure}[!h]
\centering\includegraphics[width=8cm]{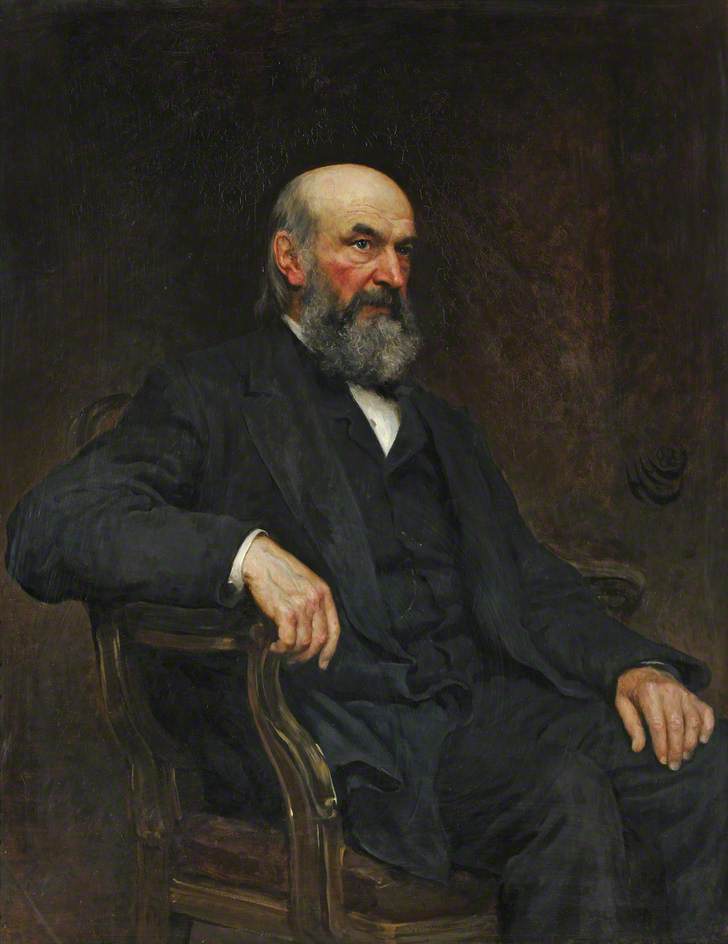}
\caption{John Couch Adams painted by Hubert von Herkomer in 1888 (aged 69). Courtesy of Pembroke College.}
\label{f:jca}
\end{figure}

To a large extent Adams' "celebrity" status was a product of nationalism, and it is questionable to what level he would be known, then or today, if behind him there was not a powerful establishment built on centuries of excellence and supporting an empire. This establishment had a use for a person like Adams. When one reads today what was written of Adams during his life within the controversies arising around the discovery of Neptune (e.g. \cite{1846MNRAS...7..121A,1846MNRAS...7..145C}), or soon after his death \cite{1892Obs....15..173G, Glaisher 1896}, one wonders which is the real Adams: the young public hero who seemed to have been let down by his peers, a Cambridge don enjoying his secluded lifestyle that allows the pursuit of interests that matter to him, or a mature Adams pointing out mistakes in works of established scientists, representing British science on the world stage and a public figure worthy of the impressive von Herkomer portraits in the National Portrait Gallery and Pembroke College (Figure~\ref{f:jca}). What is unquestionable is that he was a brilliant mathematician.

This conference is about Stokes and his legacy. As Adams and Stokes were close friends who shared in being Fellows of Pembroke College, worked together on Newton's papers, communicated about physics problems, and must have discussed each other's scientific work as well as their teaching duties, this is a good occasion to also look at Adams' achievements. After all, two centuries have passed after his birth, and 127 years after his death, so, which of his works and actions actually remain significant today? I will claim that one of Adams' most important achievements should be looked somewhere beyond the pure science, but with still relevant consequences.

\section{A short biography of John Couch Adams (1819 - 1892)}

At the time of his death, Adams was famous enough that a lot of details about his work and life found their way into literature. The most notable are the texts of James Withbread Lee Glaisher, who knew Adams personally, and wrote obituaries and a Biographical Note in Adams' collected works \cite{1892Obs....15..173G, Glaisher 1896}. Examples of further biographical details can be found in \cite{Harrison 1994, Smith 1989, Chapman 1998, SheehanThurber 2007, Hutchins2004, 2013AntAs...7...71S}. Any substantial work on the discovery of Neptune presents a number of biographical notes on Adams (e.g. \cite{1946Natur.158..648S, Jones 1947, Grosser 1962, Standage 2000, Sheehan et al. 2004}), including the latest compilation of contributions by various authors in {\it "Neptune: From Controversial Discovery to Reigning Giant of the Outer Solar System"} \cite{Sheehan2020}. Here we summarise what was written in these works and what we know about Adams' life.

\subsection{Cornwall and early studies}
\label{ss:cornwall}

John Couch, born on 5 June 1819, was the oldest child of Tabitha Knill Grylls and Thomas Adams, who lived on Lidcot Farm in Laneast near Launceston in Cornwall. They were tenants of John King Rennall Lethbridge, had a large family of seven children, and Adams' biographers often mention their relative poverty. Nevertheless, the family was able to afford a primary education for John locally at Laneast where the young lad showed his talent for algebra. An additional source of knowledge was a library of eighteenth century books, which Adams' mother Tabitha inherited from her aunt Grace and uncle John Couch. The continuation of his education as well as a decent preparation for higher education at a university happened at Devonport, where his uncle The Rev John Couch Grylls had established a preparatory school. Young Adams was interested in observational astronomy, celestial phenomena, as well as their predictions. This was the first time Adams left his family but he remained in frequent contact with them as some surviving letters testify. In one of them he wrote to his parents about observing Halley's Comet in 1835, and in another from 1836 to his brother Thomas, about the details of an upcoming solar eclipse, providing calculations on its timings and observability from their farm. In 1837, for the first time, Adams got a small share of fame. He observed a lunar eclipse with a small telescope and wrote a short summary for a journalist friend at a local newspaper. The editor of the paper later told Adams that his text on the eclipse \textit{"has been copied into several of the London papers"} \cite{Glaisher 1896}. 

\subsection{Undergraduate at Cambridge}
\label{ss:cam1}

In October 1839 Adams entered St. John's College, Cambridge, as a sizar, essentially receiving a stipend from the college to cover his costs. This college was regarded as particularly good for studies of mathematics and this might have been the reason why Adams chose it. The University of Cambridge curriculum was heavily weighted towards the mathematical subjects, and all students, including those with little interest in or inclination for mathematics, had to pass a final exam called the Mathematical Tripos. The exam lasted six days divided in 33 hours, where 24.5 hours were allocated for answering questions that appeared in prescribed books, and the remaining time was given for problems set by the fellows of the University. The award for successfully solving the problems was proportional to the difficulty of the tripos. The students taking the Tripos were ranked by their achievements and sorted in three groups called "Wranglers", "Senior Optimes" and "Junior Optimes", which would today rank as first-, second- and third-class honours \cite{Craik2007}. The top of the list was the "Senior Wrangler", followed by the "Second Wrangler " then "Third Wrangler" and so on. At the very bottom of the list, the last of Junior Optimes was called the "wooden spoon". The results of the Tripos were reported by the local and national newspapers, and the Senior Wrangler was little less than a celebrity, being paraded through the streets of Cambridge, with bettors earning or losing money. Needless to say, the future of the top Wranglers in Adams' time was secured. Often they were awarded fellowships at one of the colleges, and a few pursued careers in science and academia as distinguished professors in Britain or her colonies. Many more later found themselves holding prominent positions in the Church of England, law, or politics, while one even become a wine merchant \cite{Craik2007}. Reading a list of Senior Wranglers of that time is like reading a book of who is who in British physics. Both actors in the Neptune storey, George Biddel Airy (1823) - Astronomer Royal - and James Challis (1825) - director of the Cambridge Observatory - were Senior Wranglers. George Gabriel Stokes was the Senior Wrangler in 1841, while William Thomson (later Lord Kelvin) and James Clark Maxwell were both Second Wranglers in 1845 and 1854, respectively. John William Strutt, also known as Lord Rayleigh, was the Senior Wrangler in 1865. 

Adams was ambitious upon entry to Cambridge University, planing to attain honours in both classics and mathematics, but in the end he focused on mathematics. Part of the problem was that doing well on the Tripos was not so much related to what lectures one followed at the University, but required a special preparation. All students wishing to perform well hired private tutors to prepare them for the examinations. Adams' first tutor was Rev. Joseph Wolley (a 3rd Wrangler in 1840), but after some time he changed to John Hymers (2nd Wrangler in 1826), one of the most renowned tutors \cite{Craik2007}. The intensive preparations paid off. Adams achieved a Tripos record in 1843. Not only was he the Senior Wrangler, but he also had twice as many points as the Second Wrangler. After achieving such results Adams also took part in a further exam and obtained the Smith's Prize. By the end of his last undergraduate term Adams was elected as a fellow of (his) St. John's College. Back at home in Cornwall, the news was transmitted with exultation. Even Lethbridge, the landlord of the Adamses, took off his hat while riding a horse and shouted "Adams forever!" \cite{2013AntAs...7...71S}. 

\subsection{A Cambridge don}

Adams' undergraduate achievements were remarkable and cannot be understated. As much as the Mathematical Tripos can be criticised as a rather cruel way to test what students learned, it offered the "best mathematical foundations for the advancement of physical science available in Britain" \cite{Smith and Wise}. Nevertheless, Cambridge University of the early 19th century was not a research institution. The University was not expected to provide support for research \cite{Johnstone 2016}, and as such it was left in the domain of the professional duty, when the main task, the undergraduate teaching is finished. From the point of view of that "culture" at Cambridge, Adams' often quoted memorandum of July 3, 1841 is remarkable: {\it "Formed a design at the beginning of this week of investigating, as soon as possible after taking my degree, the irregularities in the motion of Uranus which are yet unaccounted for, in order to find whether they may be attributed to the action of an undiscovered planet beyond it; and, if possible, thence to determine the elements of its orbit \&c. approximately, which would probably lead to its discovery"} \cite{Glaisher 1896}. The background to this "research plan" is that during his undergraduate years Adams' discovered the {\it "Report on the Progress of Astronomy during the present century"}, published in 1832 \cite{Airy 1832}, which described the problem with the irregular motion of Uranus. The report, however, did not mention that the solution might be in a perturbing planet. Adams, as was later reported \cite{Somerville 1874}, might have also been drawn to the idea by reading {\it "The Connexion of the Physical Sciences"} by Mary Somerville, which, from the third edition in 1836 included an explicit mention of an external planet as a possible solution  \cite{Somerville 1836}. Regardless of the cause, Adams was focused on the task of passing the Tripos, but at the first free moment, during the long (summer) vacation in 1843, just after becoming a fellow of St. John's, he indeed turned his attention to the problem of Uranus' motion. Adams' approach and results, as well the controversy with French astronomers will be described in Section 3.a).

In spite of the failure to secure the prediction and the discovery of Neptune for Cambridge, Adams was largely perceived as the successor of Newton. He was offered a knighthood in 1847, but declined it as he could not keep up the appearance on his income \cite{1892Obs....15..173G,Hutchins2004}. In the same year the University of Cambridge established the Adams' prize in Mathematics that is still awarded every year. In 1848 he received the Copley Medal of the Royal Society and in 1849 became a member (FRS). When his fellowship in St. John's College expired in 1852, because Adams did not take holy orders, he was offered a fellowship in Pembroke College and became a neighbour of Stokes. This was an important moment in his life that would have at least two lasting consequences about one decade later, and started a life long, intimate friendship. At the same time Adams was elected as the president of the Royal Astronomical Society (1851-1853); he was president again for the years 1874-1876.

\subsection{An interlude in St. Andrews and return to Cambridge}

The 1850s in Adams' scientific life was devoted to the problem of lunar motion. At the start of the decade Adams was tutoring mathematics in his college, and started working on gravitational three body interactions, with a direct application to the system of the Sun, Earth and the Moon. In 1853 he published a paper that, I believe, should be considered his most influential scientific contribution, certainly to the field of astronomy and celestial mechanics \cite{1853MNRAS..14...59A}. In this paper he discusses the secular acceleration of the Moon's mean motion, and points out mistakes that had been made previously by Giovanni Antonio Amadeo Plana and Pierre-Simon Laplace. Adams furthermore derived his own theoretical value, which was about half that determined by the observations, but did not discuss the possible reasons for the discrepancy. 

Remarkably, Adams' results on the lunar theory initiated another controversy with French astronomers (see Section 3.b). As the controversy slowly developed on the continent, Adams was offered a chance to advance in his career. In 1857 he was elected as the Regius Professor of Mathematics at the University of St. Andrews, the only chair in mathematics appointed by the Crown. As well as the long journey north, this post required that he swear the fealty to Queen Victoria and abjure the Jacobites and Catholics. He was the last professor to have to do so \cite{Craik2007}. His stay in Scotland was, however, short. In 1859 he was already back in Cambridge, this time as the Lowndean Professor of Astronomy and Geometry, and again a fellow of Pembroke college, together with Stokes.  A general opinion by Adams' biographers is that he was "distinctly un-careerist" \cite{Chapman 1998}, but this might not be completely true. After all, a professorship lured him north of the border and another swiftly brought him back to Cambridge. In 1853 he had applied to take over the superintendence of the Nautical Almanac office, but his lack of organisational skills convinced the Admiralty to select John Hind instead \cite{Chapman 1998}.

Facing Challis' retirement as the director of the Cambridge observatory, a search was opened for a suitable candidate. Stokes played a critical role in convincing Adams to take the post \cite{Harrison 1994, Hutchins2004}. Airy was able to secure more funds from Anne Sheepshanks \cite{Airy 1896, Hutchins2004} for the observatory and Adams was offered $\pounds 250$ for the superintendence and accommodation in the "Director's Wing" at the observatory building, a classical structure with Doric style portico based on the Temple of Minerva at Athens, built by John Clement Meade in 1824 \cite{Hutchins 2008}. After several long conversations with Challis and Stokes, Adams accepted the directorship, under an understanding that he did not need to do any observations, did not have to reduce the observations, nor process the publications, and if the duty proved to be interfering with his research, he could resign \cite{Hutchins2004}. Such extraordinary terms, likely unique for an observatory director at the time, were possible as the extra money Airy obtained allowed Adams to employ an experienced assistant. 

Adams' choice of accepting the directorship had also a lasting consequence for his private life, specifically its romantic side. This episode established another remarkable connection with the family of his friend Stokes. Becoming the director of the observatory implied he would move out from the college and live at the observatory. A number of people, from the members of his family to fellow professors, advised Adams to marry \cite{Harrison 1994}. His brother William Grylls Adams feared Adams would be lonely and spend most of the time at the college instead at the observatory, and he suggested their sister, Mary Ann, move to Cambridge. The events of late 1862 allowed Adams to find {\it "the one solution of a problem, which I had well nigh come to look upon as insoluble"}, as he wrote to a fellow professor Adam Sedgwick \cite{Adams 1862}. In October 1862, through his social activities with Stokes' family, Adams met Elizabeth (Eliza) Bruce (1827-1919), an Irish friend of Stokes' wife Mary. Adams' diary for 1862 has only entries for one week in October, between 18th and 26th. On 19 October (a Sunday) Adams writes {\it "At Sr. Mary's Howson gave us a capital sermon on the conscientiousness and integrity of St. Paul. Afterwards spoke to Miss Bruce."} And then on Wedensday 22 October: {\it "Called at Porter's to get 'Rossetti's early Italian Poets'. In the evening want to Ray at Miller's. Took Rossetti to Miss Bruce."} It was a remarkable week in October 1862 \cite{Harrison 1994}.

Mary Stokes was a daughter of Thomas Romney Robinson, the director of the Armagh Observatory. It is likely that Robinson invited Adams to visit Ireland and meet Andrew Graham, director of the Markree Observatory. Adams was looking for a senior assistant and Graham was recommended to him. In December 1862 Adams went to Ireland and stayed at Armagh observatory. At the same time, Eliza also happened to be a guest there, and solved Admas' "insoluble problem" he had referred to in the letter to Sedgwick. They were married on 2 May 1863 at Bray, Co. Wicklow. Their marriage was, by all accounts, a happy one with shared interests including the occult and mesmerism \cite{Harrison 1994}, and with Eliza likely taking care of the managerial aspects of their life. In preserved letters between Eliza and Stokes one can see that she supervised their finances and investments (Jayne Ringrose, private communication). At the end of his life, Adams' wealth was declared to be close to $\pounds 32500$ \cite{Hutchins2004}. 

The visit to Ireland was a great success for Adams. Next to a bride, he also hired a capable astronomer, Graham, who essentially led the work of Cambridge observatory during Adams' tenure between 1862 and 1892. Graham's expertise was in cataloguing and charting accurate star positions, for which he ordered a transit circle in 1867 and which became operational in 1870. This coincided well with a request from the Astronomische Gesellschaft to observe one of the last zones in their charting of the sky, which started in 1860. This was a natural element for Graham as the region in question, between 25 and 30 degrees north, overlapped with his previous work. The work started in 1872 and was completed in 1886, but published only in 1897, after Adams' death. Adams' directorship of the observatory seemed to have been rather easy going, without exerting a pressure on his staff to publish their work, with observations between 1861-1865 being published in 1879 and those of 1866-1869 only in 1890 \cite{Hutchins2004}. 

The directorship also brought further public duties to Adams. He was elected to the board of Visitors to the Royal Greenwich Observatory and he continued to be a member of the Royal Astronomical Society Council and of the RAS Club. He was the president of the Cambridge Philosophical Society (1874-1876). Such positions naturally bring about a certain level of influence. Still, Adams tried to avoid all controversies and to stay out of any kind of politicking. On the other hand, he was generous to fellow astronomers. He proposed Hencke and Hind for the 1848 RAS Gold Medal, and in 1879 even opposed Airy to ensure that Hind was elected the president of the RAS. When he was the president of the RAS for the second time, he had the good fortune to deliver the addresses on the presentation of the Gold Medals to Hind \cite{1853MNRAS..13..141A}, and during his second term to d'Arrest \cite{Adams 1875} and Le Verrier \cite{Adams 1876}. Ironically, he did not know, like most of the world at the time, that d'Arrest was one of the discoverers of Neptune, otherwise this would certainly be mentioned in his address. In 1870, acting as a Vice-President of the RAS he also had the fortune to deliver the Gold Medal address for Delaunay \cite{Adams 1870}, his French champion in the Lunar Theory controversy (see Section~3.b).

In 1881 Adams was offered the position of Astronomer Royal, upon Airy's retirement. He refused it as he considered himself to be too old for the job \cite{Craik2007}. In 1884 he travelled to Washington as part of the British delegation to the International Prime Meridian Conference. He was certainly there as a director of a major British observatory, and he made clear contributions to the heated discussions \cite{Harrison 1994}. Nevertheless, it is curious that the British delegation did not include the Astronomer Royal. It is hard to think of any other reasons for his attendance other than as a famous predictor of Neptune, the successor of Newton and the symbol of the brilliance of the British scientific establishment. 

In October 1889 Adams became seriously ill. He recovered sufficiently to resume his mathematical work in early 1890, but the illness returned several times over the next year and a half. Adams died on 21 January 1892 and was buried in St. Giles cemetery in Cambridge. 

\section{Adams' scientific achievements}

\subsection{The Uranus problem} 
\label{ss:uranus}

William Herschel, while conducting the Bath Orchestra during the day and counting double stars during the night, discovered Uranus in 1781. Soon after, it became clear that Uranus had been mistaken for a star and observed many times previously, as early as 1690 by the Astronomer Royal John Flamsteed. This allowed for a fairly precise calculation of the 84 year orbit of Uranus. However, the orbit based on the pre-discovery observations could not be reconciled with subsequent observations \cite{Bouvard 1821}. The difference between the predicted and observed locations of Uranus, once the influence of all major Solar System bodies was taken into account (Jupiter and Saturn), could be as high as 60 seconds of arc, where the observational accuracy was considered to be better than 5 seconds of arc. This was a major issue for the astronomical establishment, and a number of observatories (Paris, Cambridge, Greenwich, Königsburg) established long lasting campaigns to track the motion of Uranus. Crucially, this was potentially a serious problem for Newton's theory of gravitation. 

Alexis Bouvard, who presented the first complete tables for Uranus, considered that the solution of the problem must be in an external planet \cite{Bouvard 1821}. Airy, who also measured the radial movements of Uranus \cite{1838AN.....15..217A}, was open to an idea that the law of gravity might be somewhat different at such vast distances from the Sun \cite{Airy 1846let}. Friedrich Wilhelm Bessel went the furthest and performed a number of tests of Newtonian gravity, including a theoretical modification of gravity, the {\it Selective Attraction} theory \cite{Hamel 1984}. His idea was that the mass with which Saturn acts upon Uranus and the mass with which Saturn acts with the Sun are not the same, resulting in a modification of Newton's law for Uranus' motion. Bessel shared some of his calculations with only close collaborators, but concluded that this idea does not work. By the mid 1830s, the general opinion was that a planet might be responsible and Bessel even put a student, Friedrich Wilhelm Flemming, to the task to derive the orbit of the planet based on the observation of Uranus. Flemming, however, died in 1840, completing only the reduction of Bessel's observations of Uranus \cite{Schumacher 1850}.

Adams' memorandum in 1841 is not the earliest attempt to solve this problem. His approximate calculation of the orbit in the summer of 1843, and the first complete solution in the early autumn of 1845 take precedence as the first successful solutions to the problem. In the first try in 1843, Adams was interested to see if the orbit of Uranus can be explained by assuming there is an external planet, where the distance and the eccentricity of the orbit are restricted by some strong assumptions: for distance Adams assumed the prediction of the Titius-Bode law\footnote{Johann Daniel Tietze showed that one could approximately represent the distances of the known planets from the Sun as a simple series: $a=0.4 + 0.3\times2^{n-1}$. In this equation, $a$ is the semi-major axis of the orbit expressed in the units of the semi-major axis of the Earth's orbit (astronomical unit), and $n$ is the integer number representing the planets in their order: $-\infty$ for Mercury, 1 for Venus, 2 for Earth and so on. Tietze, also known under his latinised name Titius, published this result as an addition to the original text in his translation of Charles Bonnet's "Contemplation de la Nature" in 1766. He might have been aware of previous attempts to sort the planetary distances into a series, for example the one by David Gregory in "The Elements of Astronomy" in 1715. Johann Elert Bode added Titius' expression to his "Anleitung zur Kenntniss des gestirnten Himmels" in 1772, which contributed to its popularity. The Titius-Bode formula was raised from the status of a curious empirical finding to a "law" when Uranus and Ceres were found to be well represented by it. The law, however, breaks down with Neptune and Kuiper belt objects.} and that the orbit was circular. His calculations showed him that he was on the right track. When he restarted the work, using better data (which he obtained in the mean time from Airy via a request by Challis), he provided a more accurate (non-circular) orbit. By September 1845 Adams had the first orbit of Uranus' perturber, which he felt confident to show to his peers. 

James Challis, with whom Adams established a collaborative relationship at that time, must have sensed this was a matter of high importance and pointed Adams to the Astronomer Royal Airy. He wrote a letter of introduction for Adams and a few days later Adams, on his way to Cornwall, passed by Greenwich. Airy was in France, and Adams simply went on to his family. During the vacation, he proceeded with the work on Uranus and made improvements on the perturber's orbit. On his way back from Cornwall, Adams again passed by Greenwich, this time twice missing the Astronomer Royal: first as Airy was outside on a walk and second as the family was at the dinner table, and Adams was not admitted. However, he left a note dated 21 October 1845 with the reworked orbit. 

The events between the summer of 1843 and the early autumn of 1846 marked Adams' life in many ways. Adams had predicted the location of the planet, but he failed to publish it. Moreover, even though he was in a direct communication with Airy about the problem, he failed to answer a question that Airy considered very important. Airy, almost uniquely among astronomers, was also interested in how much the predicted radial component of the Uranus orbit differed from the observations. Determining the precise distance to Uranus along its orbit was a much harder observational task than the longitudinal positions pursued by everybody else, but Airy published it in 1838 \cite{1838AN.....15..217A}. He concluded that the "radius vector" of Uranus is also at odds with the theoretical predictions based on Bouvard's tables. When Airy saw Adams' prediction for the orbit of the new planet, he considered the question of the radius vector an "experimentum crucis" \cite{Airy1846}: Adams' theory had to explain the error in the distance to Uranus before Airy could accept it as sound.

Adams did start drafting an answer to Airy's query in November 1845, but never finished it \cite{Sheehan 2020b}. This had the consequence of Airy dropping the matter and the problem of Uranus got almost forgotten in Britain. It was different in France, where Francois Arago, the main figure of the French astronomical establishment, asked Jean Joseph Urbain Le Verrier to tackle the problem of Uranus. Le Verrier was not a trained astronomer, he started his scientific career as a chemist. As he was able to get a teaching job in astronomy at the  \'Ecole Polytechnique, Le Verrier changed subjects and presented himself with a few well received papers as a capable "mathematical astronomer". Le Verrier announced in November 1845 his analysis of Bouvard's tables and concluded that a planet must be responsible for the perturbations \cite{Le Verrier 1845}. In June 1846 he presented his first solution, predicting the orbit and the location of the perturber on the sky \cite{Le Verrier 1846a}. 

At the same time, Adams' main activity was teaching, and even during the vacation time he was preparing himself for the tutoring sessions \cite{Sheehan 2020b}. The non-competitive environment of Cambridge University, at least for research, probably did not help either. He seems to have been aware of the prize announced by the Academy of Göttingen for the problem of Uranus in 1844, but this did not push him to publish his own result \cite{1846MNRAS...7..149A}. When it became clear that Le Verrier had also reached the same prediction on the location of the perturbing planet, Airy pushed Challis to start the search for the planet \cite{1846MNRAS...7..121A}. Adams was involved in this project from the start by calculating ephemerides based on possible orbits (starting from the one of Le Verrier). Nevertheless, the Cambridge search failed; the new planet was discovered in Berlin on 23 September 1846, by Johan Gottfried Galle and Heinrich Louis d'Arrest, based on the prediction of Jean Joseph Urbain Le Verrier\footnote{Le Verrier wrote directly to Galle asking him to search for the planet. He also wrote to Otto Struve at the Pulkovo Observatory, but that letter came too late: the letter of the discovery from Berlin reached Pulkovo before Struve was able to start observing \cite{Dick 1986}.} \cite{Le Verrier 1846b}. The news reached England by a letter sent from Franz Friedrich Ernst Br\"unow to John Russel Hind (dated 25 September 1846), who after observing the new planet sent the announcement to The Times dated 30 September 1846. 

In October 1846 Adams found himself in the centre of both a British and an international scandal. It all erupted when Sir John Herschel announced on 3 October, in a British literary magazine The Athenaeum, that {\it "a young Cambridge mathematician, Mr. Adams"} had also predicted the location of the new planet \cite{Herschel 1846}. Challis published on 15 October, first in The Athenaeum and then in the Astronomische Nachrichten \cite{Challis 1846a, Challis 1846b}, the description of his search for the planet. This announcement was not only astonishing because of the existence of the Cambridge search itself, but also because it turned out that Challis had observed the planet already at the beginning of August, but failed to analyse the data in time to actually discover it. Nevertheless, this observation gave another point based on which Adams was able to calculate the new orbit, and, specifically, determine the distance to the new planet. This was another surprise, as the distance turned out to be rather different as compared with the predictions of both Le Verrier and Adams. This later led Benjamin Pierce, Perkins Professor of Mathematics and Astronomy at Harvard, to question if the discovered planet is the same as that predicted. Pierce stated it was a "happy accident" the planet was discovered based on the prediction, either Le Verrier's or Adams' \cite{1992JHA....23..261H}.

The international part of the scandal revolved around two issues. It was difficult for the French establishment to accept that {\it after the discovery} there is a legitimate claim for a co-prediction, and that the glory should be shared with an essentially unknown Cambridge mathematician. For Le Verrier personally, this was even more difficult to understand as he was already in communication with Airy in June 1846. Airy started the correspondence on 26 June asking Le Verrier the same question about Uranus' radius vector that Adams had failed to answer. Le Verrier sent the answer on 28 June confirming that the error in the radius vector was accounted for, explaining the details of his theory, but also asked {\it "If I could hope that you will have enough confidence in my work to seek this planet in the sky, I would hasten, Sir, to send the exact position to you, as soon as I have obtained it."} \cite{Le Verrier to Airy 1846}. Airy did not ask Le Verrier for a better prediction as he was able to see that Le Verrier's prediction was within a few degrees from that of Adams. Le Verrier's estimate of the error on the predicted location at that time was 10 degrees, but the fact that two people reached essentially the same conclusion independently, using somewhat different data and differing in the approach (see \cite{Baghdady 1980} and \cite{1970CeMec...3...67B} for details on the mathematical approaches), convinced Airy that the search should start as soon as possible (see also \cite{2015JHA....46...48G} about the precision of Le Verrier's final prediction). 

The other part of the international scandal relates to the story of naming the planet. This is by far the most illustrative and entertaining part of the story (see \cite{Wattenberg 1962, 2009JAHH...12...66K, Krajnovic 2020}), and Challis and Adams can be blamed for inflaming it. Galle in his letter to Le Verrier announcing the discovery suggested the new planet be called Janus, being so far out of the Solar System. Le Verrier rejected it by saying that this would imply the planet was the last in the system, and wrote that the Bureau des Longitudes had chosen the name Neptune (although this seems to actually be his own invention). This exchange between Galle and Le Verrier was circulated in the newspapers and in Britain it was transmitted by The Athenaeum on 10 October 1846. In spite of that Challis wrote in his article of 17 October: {\it "The part taken by Mr. Adams in the theoretical search after this planet will, perhaps, be considered to justify the suggesting of a name. With his consent, I mention Oceanus as one which may possibly receive the votes of astronomers"} \cite{Challis 1846a}.

The national part of the scandal was all about {\it "depriving this country and his own University of the merit of the first announcement"} \cite{Glaisher 1896}. The British public and scientific establishment was shocked by the fact that Adams' prediction was about 9 months before Le Verrier's and that nothing was published about it. The blame was put at Challis and Airy for not giving {\it "the young and retired man the kind of help or advice that he should have received"} \cite{Glaisher 1896}. Writing a few years after Airy's death, Glaisher explicitly puts the blame on the Astronomer Royal for not making sure that Adams' work was published. Such opinions remained well over hundred years in the literature about the discovery of Neptune (e.g.\cite{Grosser 1962, Standage 2000}), but recent studies are more sympathetic to Airy (e.g. \cite{1988JHA....19..121C, Sheehan2020}) and Challis \cite{2013AntAs...7...71S}. 

Was Adams an inexperienced youth who had a {\it "difficulty, which all young writers feel more or less, in putting into shape and order what he has done"} \cite{Challis to Airy 1846}? Or did Adams act {\it "like a bashful boy rather than like a man who had made a great discovery"} \cite{Sedgwick to Airy 1846}? This remains entirely a matter of opinion. The facts are that Adams had made an astonishing prediction, but failed to convince his peers that it was a serious theory. The moment Airy saw a confirmation of Adams' prediction, he acted and Challis started a serious observational campaign. His approach was  perhaps not very enthusiastic, but it was professional. He might have taken a sky chart made by Argelander for the Berlin Academy in 1832 (distributed in 1833) which covered the region where Neptune was at the end of July (at the beginning of his search), and would most likely have seen it within a night, as it happened in Berlin\footnote{Berlin observers, looking for the planet almost 2 months later, used another chart made by Carl Bremiker in 1845. The chart lay forgotten in the director's office, waiting to be distributed together with another still unfinished chart. The fact that the map was only present in Berlin gave rise to various speculations about the importance of this map for the discovery, ranging from the idea that Galle and d'Arrest used it from the start of the search to the idea that Le Verrier knew about the map in the first place. Its absence from Cambridge was often assumed as the crucial reason the planet was not discovered by Challis \cite{Dreyer 1882}. The truth is that the Berlin observers proceeded with the then standard Harding's Sky Atlas and were looking for a planet based on Le Verrier's prediction that it would be seen as a disk. Only when they were not successful, they thought it would be good to check if any of the new maps covered the area of the search \cite{Galle 1877, Galle 1882}. They were lucky.}. His approach was different, however, attempting to chart the stars himself in the most rigorous way, but also continuing to work on other projects that interested him more (i.e. comets). In the end, it is important to state that Adams was not drawn into any of the discussions and accusations around him. He never said a word against Challis, or Airy, or Le Verrier. 

\subsection{The lunar theory}
\label{ss:moon}

The secular acceleration of the Moon's mean motion refers to the fact that the the Moon is gradually going faster in its orbit around the Earth. This was first pointed out by Edmond Halley \cite{Halley 1695} based on the comparison of the duration of ancient ellipses. The first numerical determination of the effect amounting to about 10 arcsec per 100 years was given by Dunthorne \cite{Dunthorne 1749} comparing the data of the lunar eclipses of ancient Babylonians in 8th, 4th and 2nd century BC, as recorded by Ptolomey, and the observations of Ibn Junis at Cairo in the 10th century. Somewhat later Tobias Mayer revised his initial value of 6.7 arcsec/century to 9 arcsec/century \cite{Mayer 1770}, and Lalande provided a value of 9.886 arcsec/century \cite{Delaunay 1863}. The explanation of this effect was considered a major open question for the Newtonian theory of gravity and it featured in a series of prize questions established by academies\footnote{The French Academy of science proposed the question in 1762, 1770, 1772 and 1774, while the Swedish Academy of Sciences proposed it in 1787 (for 1791) and awarded the prize to Laplace. Winners of the French competitions were Bossut, Euler, Euler and Lagrange, and Lagrange, respectively. }. The most satisfactory solution was found by Laplace in 1787, who showed that the observed effect can be explained by considering the variation in the eccentricity of the Earth's orbit due to the influence of other planets. As Earth's orbit becomes more circular, the mean distance of the Earth from the Sun increases, and the perturbative force of the Sun on the Moon, therefore, decreases. This has the effect that the net acceleration towards the Earth is increased and the Moon's motion is accelerated.

In 1853, Adams published a paper \cite{1853MNRAS..14...59A} as a reaction to the calculation that Plana (and also published separately by Marie-Charles-Théodore de Damoiseau) performed for the Grand Prize of Mathematics in 1820, set by the French Academy of Sciences with a general topic of the lunar theory. Plana extended the work of Laplace by expanding the perturbation to 28 terms (Laplace's result was based on the first term in the series), and showed that the expected variation is 10.58 arcsec/century (Damoiseau obtained 10.78 arcsec/century using a different method) \cite{1989AHES...39..291K}. However, Adams noted that Plana made an error in assuming that the Earth's eccentricity is constant within an orbit, while in reality it is continually diminishing. Once this is taken into account, Adams obtained the theoretical value for the secular acceleration of the Moon of 5.70 arcsec/century In 1857, Peter Andreas Hansen published {\it "Tables de la Lune"} \cite{Hansen 1857} with an observational value of 12.18 arcsec/century. 

Adams' paper, contradicting both the observed value and the work of Laplace and Plana, but not providing any explanation for the origin of the rest of the effect, did not cause much of a reaction immediately. However, it soon developed into a full blown controversy \cite{1989AHES...39..291K}. First it was Plana who reacted in 1856 and claimed that Adams was correct and then that he was not, providing another solution (as referenced in \cite{Delaunay 1863}). In 1859 Delaunay \cite{Delaunay 1859a} published a paper which confirmed Adams' result as well as extending the work to larger terms (Adams communicated his own value of these higher terms to Delaunay in a letter and Delaunay announced them in his own paper \cite{Delaunay 1859b}), and reached the result of 6.11 arcsec/century. Philippe Gustave Doulcet de Pont\'ecoulant joined the discussion \cite{Pontecoulant 1859a, Pontecoulant 1859b} by claiming that Adams introduced terms that were rightly ignored by Laplace, Damoiseau, Plana and himself. The controversy was deepened by Le Verrier \cite{LeVerrier 1860a,LeVerrier 1860b,LeVerrier 1860c}, whose point was that Delaunay's calculation (and Adams' as well) must be flawed as it does not correspond with the observations. 

It is interesting that Le Verrier did not actually mention Adams, but attacked directly Delaunay, even though Delaunay was essentially defending Adams' results. This could be interpreted in many ways. Le Verrier might have wanted to show indirectly that Adams, the person who "spoiled" his Neptune prediction, was wrong, continuing a sort of scientific vendetta \cite{1989AHES...39..291K}. On the other hand, the fact that Le Verrier does not mention Adams at all could be considered in an opposite sense, that Le Verrier did not want to draw Adams into a long standing fight he had with Delaunay (for details see \cite{Lequeux 2013}). Adams was most directly attacked by Pont\'ecoulant, both in French and British journals \cite{Pontecoulant 1859a,Pontecoulant 1859b}, and a biographer of Adams should not forget his rebuttal of the accusations published in the Monthly Notices in 1860. The Adams of 1860 was a very different person from the Adams of the mid 1840s. Now he was able to write: {\it "Again, I had some hopes that M. de  Pont\'ecoulant might be led to see and acknowledge the errors into which he had fallen, and with that object in view I sent him, on more than one occasion, through a friend, communications which appeared to me amply sufficient to expose the fallacies contained not only in his printed "Observations", but also in several private letters which he subsequently wrote upon the subject"} \cite{1860MNRAS..20..225A}. 

The controversy erupting from Adams' work slowly disappeared, as it became clear that Delaunay's and Adams' work could not be challenged mathematically \cite{Delaunay 1867}. The question of what is the cause of the other half of the effect remained unanswered for a few more years. At the time the Royal Astronomical Society awarded its Gold Medal to Adams for his lunar theory work in 1867, the other cause of the effect started to crystallise. In 1863 Hansen pointed to the likely source of the effect: shortening of the rotation period of the Earth by 0.01197 seconds in 2000 years \cite{1863MNRAS..23..210H}, and Delaunay in 1865 \cite{Delaunay 1865} worked out that this can be achieved by the Moon's influence on the Earth's tides, introducing the concept of tidal retardation in the lunar theory\footnote{The priority should actually be given to William Ferrel \cite{Ferrel 1866} who presented the same result to the American Academy of Arts and Sciences in December 1864, but his paper was not published until 1866.}.

Adams continued to work on the lunar theory until the end of his career. In 1877 he communicated part of his work to the Monthly Notices on the motion of the lunar perigee. This was motivated by a publication of the same result by George William Hill, which he acknowledged and then sketched his own solutions to the problem. In 1880, after Airy presented a paper on the secular acceleration of the Moon's mean motions, recovering Laplace's results, Adams had no scruples and pointed out mistakes Airy made in his calculation. Subsequently, Airy corrected his work and concluded with result that differed little from Adams' 1853 results. 

\subsection{The Leonids and other works}
\label{ss:other}

Adams' third widely known scientific contribution is that of determining the origin of the November meteor shower, or Leonids as they are called now. This was taken on by him as another interesting problem in celestial mechanics following the prediction of Huber Anson Newton when a meteor shower would occur in 1866 from a radiant point in Leo. The remaining question was to show the exact orbit of the shower, given that: i) it is possible to observationally determine the radiant point of the meteor shower, ii) that Newton determined for how much the longitude of the node of the orbit of the meteors is increasing annually, and iii) that there were five possible orbital periods. By March 1867, Adams was able to show that only one orbit was consistent with the observations, and that it had a period of 33.25 years. Adams included the effects of Jupiter, Saturn and Uranus, which were found to be responsible for the increase of the longitude of the node. In this way, Adams established the link of the meteor shower with a comet discovered by Wilhelm Tempel in 1865 \cite{1867MNRAS..27..247A}. As Adams acknowledges in the same paper, Giovanni Schiaparelli on 31 December 1866 and Le Verrier on 21 January 1867 published their calculations with similar orbits for the meteor shower and the same conclusion for its cometary origin. 

This example is indicative of the full publishing career of Adams. In the Obituary \cite{1892Obs....15..173G} Glaisher writes: {\it "It is also noticable that so few of his papers should have appeared quite spontaneously: it frequently happened that he was incited to give an account of something which he had done himself - probably years before - by the publication of a paper in which the same ground was partially covered by some other investigator; in other cases he was called upon to correct some misapprehension which was leading others astray"}. Adams seems to have been an idealist, choosing topics of his research because they interested him and working on them while that interest lasted. He could also be considered a perfectionist, who postponed publication in order to make further improvement and attain more precise results. The Neptune affair, some parts of the lunar theory, and the Leonids are examples of such behaviour. Only the 1853 paper on the secular acceleration of the Moon's mean motion can be considered truly influential, in the sense that it {\it motivated} further research and eventually the explanation of the phenomenon. 

Adams' contribution to pure mathematics was small, but very curious. He worked out the Bernoulli numbers from 31 to 62, and then used them to calculate Euler's constant to 263 decimal places. In order to do this he had to derive the logarithms of 2, 3, 5 and 7 to the same precision, which he later extended to ten more decimal places. If nothing else, these works testify to the "patience and skill" he devoted to his research \cite{Adams 1896}. 

As mentioned several times, Adams was often compared with Newton. His contemporaries at the university certainly considered him as such \cite{Glaisher 1896}. Therefore it is not surprising that he was put in charge, together with Stokes, of sorting out Newton's unpublished papers that were presented to the University by Isaac Newton Wallop, Lord Portsmouth. This seems to have been the only actual collaboration between Adams and Stokes, although there is evidence that Adams was asked for opinions on problems related to ballistics and projectiles by the Ministry of Mines under Stokes \cite{Harrison 1994}. Both Stokes and Adams were very well suited to distill Newton's scientific legacy and they focused on the scientific papers covering physics and mathematics provided by Lord Portsmouth. Their work finished in 1888 with a publication of a catalogue of the papers, where the mathematical part was analysed by Adams \cite{Glaisher 1896}.

\section{Adams' Cambridge legacy}

Adams' legacy is a problematic topic, both in terms of him as a scientist and as a public person. He predicted the location of a planet perturbing the motion of Uranus, but this prediction did not help in the discovery. Neptune was discovered in Berlin based on the prediction of Le Verrier. Adams' work was a remarkable confirmation of Newtonian gravity, but this theory was superseded by general relativity. He pointed out the problem in the understanding of the secular acceleration of the Moon's mean motion, but he didn't explain the effect with a new theory. This was done by Delaunay who took over the lead in the field. His explanation of the Leonids coincided with that of other astronomers (e.g. Schiaparelli, Le Verrier). Is there a lasting legacy of Adams? One that we can today compare with that of, for example, Stokes?

I would like to argue that, although Adams' scientific legacy is not comparable to that of Stokes, he was ahead of his time for something that still resonates today, and has not ceased to be any less relevant. From 1869 Adams was an active supporter for the higher education of women in Cambridge \cite{Hutchins2004}. He was rare among professors in allowing women to attend his lectures \cite{Craik2007}. He was the first president of the Association of Promoting the Higher Education for Women in Cambridge and he was involved in establishing Newnham College (for women). One could perhaps argue that Adams' most significant decision as the director of the Cambridge observatory was to employ a "Lady Computer" in 1879, eleven years before the Greenwich Observatory \cite{1995QJRAS..36...83B,Mullen 2020}.

Anne Walker was 15 years old when Adams employed her as an assistant to Graham. In 1885 she was referred to as "a good observer and an expert calculator", and her salary increased from $\pounds 40$ to $\pounds 100$ per year by 1895 (including free accommodation at the observatory), at that time the highest paid female in British astronomy \cite{Chapman 1998}. She was not the first female employed by the observatory, as at least 5 others worked between 1876 to 1904, usually for one or two years \cite{1998AsNow..12...48B, Hutchins 2008}. However, Walker was not only a "computer" used to analyse the data obtained during the nights, but actively participated in observations, together with Graham, and later on her own. Adams must have recognised this as he was increasing her salary by $\pounds 10$ every three years from 1884. This is notable as Graham was on the same salary increase scheme of $\pounds 10$ per three years (although his starting salary was $\pounds 150$ \cite{Hutchins 2004b}). 

After Adams' death things changed for Walker. The next director of the observatory, Robert Stawell Ball did not know what to do with her, and looked for external candidates for the vacant position of the 2nd assistant, hiring a 23 year old Trinity College graduate \cite{Hutchins 2004b}. Ball also stopped Walker's regular salary increase, even though she was crucial in helping Graham (who was then in his late seventies) to finish the observations of the zone for the Astronomische Gesellschaft \cite{Hutchins 2008}. When Graham retired in 1903 (aged 88), Walker resigned on the same day. Anne Walker worked at the observatory for 24 years, but left without a pension and emigrated to Australia, leaving no trace at the observatory except several formal notes in reports and catalogues \cite{Hutchins 2004b}.

Anne Walker, together with Caroline Herschel and Mary Edwards, was a pioneer among British female astronomers. Caroline Herschel was the first known woman to pursue observational astronomy in Britain \cite{Bruck 2009}, and she also received an annual stipend from King George III for assisting her brother. Mary Edwards performed calculations for the {\it Nautical Almanac} for more than 37 years, actually working from home, and at a later point with her daughters, in a special arrangement with the Astronomer Royal Nevil Maskelyne \cite{Mullen 2020}. In that respect, Anne Walker is not the first woman to be hired as a calculator, but she swiftly evolved to become an assistant and an observer. The way Adams employed Anne Walker at Cambridge Observatory could almost be considered as a precedent for the hiring of women "computers" at the Greenwich Observatory by the Astronomer Royal William Christie in 1890s, who were much more than just "computers" and pursued observational careers \cite{Mullen 2020}. There is, however, no direct evidence that Christie got the idea from Adams, or that he was aware of Anne Walker. One could perhaps even draw a parallel between Adams' scientific life and his role as an advocate for female astronomers: he might have been the first to tackle the problem, but others pursued it further.

Adams' move to employ women was most likely initiated by the same reasoning that Edward Charles Pickering used when he started employing female calculators at Harvard from 1881; or why university educated women started working at the Greenwich Observatory in 1890. They were highly educated, but cheap (e.g. \cite{1995QJRAS..36...83B,Chapman 1998}). To cut costs, observatories were employing boys and kept them in an apprenticeship system. "Lady computers" were older and already educated, therefore removing the need for supervision and teaching of the basics. Anne Walker is, therefore, an exception as she was only 15, of similar age to the boys that were typically sought after by observatories. Crucially, Anne Walker was a female talent that was spotted, nurtured and supported to work on observational astronomy. This is the lasting legacy of John Couch Adams.


\enlargethispage{20pt}




\competing{The author declares that he has no competing interests.}


\ack{I would like to thank the organisers for an enlightening conference and Pembroke College for wonderful hospitality. This work would not be possible without the excellent support of Regina von Berlepsch and Melissa Thies, the librarians of the AIP. I also thank Barry Rothberg and Julyan Cartwright for proofreading and commenting on the manuscript.}



\end{document}